\documentclass[12pt]{article}
\usepackage{wrapfig}
\usepackage{epsfig}
\usepackage{hyperref}
\usepackage{amssymb}
\usepackage{amsmath,mathtools}
\usepackage{amsfonts}
\usepackage{latexsym}
\usepackage{wasysym}
\usepackage{multirow}
\usepackage{fixmath}
\usepackage{color}
\usepackage{stackrel}
\usepackage{txfonts}                                                            \usepackage{centernot}
\usepackage{slashed}
\usepackage{bbm}
\usepackage{bbold}
\usepackage[low-sup]{subdepth}
\usepackage{hyperref}

\newcommand{\bea}{\begin{eqnarray}}
\newcommand{\eea}{\end{eqnarray}}
\newcommand{\bite}{\begin{itemize}}
\newcommand{\eite}{\end{itemize}}

\newcommand{\MSbar}{\,\overline{\!\rm MS\!}\;}
\date{}

\begin{document}
\title{
\vspace{-1.75cm} 
\flushleft
{\normalsize DESY 25-124} \\
%\vspace{-0.35cm}
%{\normalsize Edinburgh 2017/04} \\
%\vspace{-0.35cm}
%{\normalsize Liverpool LTH 1122} \\
%\vspace{-0.35cm}
%{\normalsize March 2017} \\
\vspace{0.75cm}
\centering{\bf Renormalization Group Approach to Confinement}\\[0.5em]}

\author{Gerrit Schierholz\\[1.25em] Deutsches Elektronen-Synchrotron DESY,\\ Notkestr. 85, 22607 Hamburg, Germany\\
  and\\
  II. Institut f\"ur Theoretische Physik, Universit\"at Hamburg\\
Luruper Chaussee 149, 22761 Hamburg, Germany}

\maketitle
%\vspace*{-0.5cm}

\begin{abstract}
  While we have several complementary models of confinement, some of which are phenomenologically appealing, we do not have the ability to calculate analytically even simple aspects of confinement, let alone have a framework to eventually prove confinement. The problem we are facing is to evolve the theory from the perturbative regime to the long distance confining regime. This is generally achieved by renormalization group transformations. With the gradient flow we now have a technique to address the problem from first principles. The primary focus is on the running coupling $\alpha_S(\mu)$, from which confinement can be concluded alone. A central point is that the gluon condensate is scale invariant, which reflects its self-similar behavior across different scales. Building on that, we derive $\alpha_S(\mu) \simeq \Lambda_S^2/\mu^2$, which evolves to the infrared fixed point $1/\alpha_S = 0$ in accordance with infrared slavery. The only important factor appears to be the presence of the gluon condensate,
  % represented by homogeneous vacuum fields,
which is a universal feature that QCD shares with many other models. The analytical results are supported by numerical simulations.

%It is in the nature of renormalization group transformations that ultimately only homogeneous vacuum fields, represented by condensates, are relevant for the problem, which is a universal feature of many other models.
  
\end{abstract}
%\vspace*{0.25cm}

%\newpage
\section{Introduction}

The most salient feature of QCD is color confinement. Confinement cannot be adequately described by perturbative methods but requires nonperturbative techniques for analysis. The nonperturbative structure of the theory manifests itself in the presence of vacuum condensates, analogous to those in condensed matter physics. They refer to the lowest energy state of the system, the ground state. Most important is the gluon condensate, which is an anti-screening, color paramagnetic medium formed by gluon self-interaction. The effect it has on a colored test charge is revealed by the renormalization group. %This paper describes what happens when the scale parameter is taken to the infrared limit. 

%The effect of virtual gluons is to increase the QCD effective charge with increasing distance. The tool that puts this effect on a theoretical basis is the renormalization group. %Also, gluons provide 99\% of the visible mass of the Universe.Its properties are crucial for understanding how gluons and quarks are confined within hadrons. 

The renormalization group approach is a strategy for dealing with problems involving different length scales. With the gradient flow~\cite{Luscher:2010iy} we now have a tool that is a continuous realization of Wilson's renormalization group transformation of the action, and as such provides an analytical means of exploring the nonperturbative properties of the theory. It evolves the gauge field along the gradient of the action. The flow of gauge fields is defined by
\begin{equation}
\partial_{\,t}\,B_\mu(t,x) = D_\nu G_{\mu\nu}(t,x) \,, \quad G_{\mu\nu} = \partial_\mu\,B_\nu -\partial_\nu\,B_\mu + [B_\mu, B_\nu] \,,
\label{gf}
\end{equation}
where $D_\mu$ is the covariant derivative and $B_\mu(t=0,x) = A_\mu(x)$ is the  initial gauge field generated by the action
\begin{equation}
  S_G = \frac{1}{2 g_0^2} \int d^4x\, \textrm{Tr}\, F_{\mu\nu} F_{\mu\nu} \,, %\,, \quad F_{\mu\nu} = \partial_\mu A_\nu - \partial_\nu A_\mu + [A_\mu, A_\nu] 
\end{equation}
where $g_0$ is the bare coupling constant, which is kept fixed. At tree level the solution at flow time $t$ is obtained by convolving the initial gauge field with a Gaussian heat kernel, $K(x,t)=\exp\{-x^2/4t\}/(4\pi\, t)^2$, of variance $\sqrt{2t}$. This means, the gauge field is averaged over a spherical range in space-time with mean-square radius $\sqrt{8t}$. A key property of the flowed field is that the tree-level result extends to all orders in perturbation theory. A natural choice for the renormalization scale $\mu$ is the inverse mean-square radius $\mu = 1/\sqrt{8t}$. As shown elsewhere, $1/\sqrt{t}$ serves as a Wilson-type ultraviolet cut-off~\cite{Beneke:2025hlg}. A notable feature is that the flowed gauge fields in correlation functions do not require further renormalization beyond the theory's bare parameters~\cite{Luscher:2011bx}.

We denote the vacuum energy density at flow time $t$ by
\begin{equation}
  E(t) = \frac{1}{2}\, \textrm{Tr}\; G_{\mu\nu} G_{\mu\nu} \,.
  \label{density}
\end{equation}
The vacuum expectation value of $E(t)$ defines a renormalized (running) coupling in the gradient flow ($GF$) scheme~\cite{Luscher:2010iy,Harlander:2016vzb},
\begin{equation}
\alpha_{GF}(\mu) =  \frac{4 \pi}{3}\, t^2 \langle E(t)  \rangle \,.
\label{coupling}
\end{equation} 
The corresponding $\beta$ function is given by
\begin{equation}
  \mu \,\frac{\partial\, \alpha_{GF}(\mu)}{\partial\, \mu} = -2t \, \frac{\partial\, \alpha_{GF}(\mu)}{\partial\, t} = \beta_{GF}(\alpha_{GF}) \,,
\end{equation}
which encodes the dependence of $\alpha_{GF}(\mu)$ on the scale parameter $\mu$. In fact, as we move down in energy $\mu$ the theory makes a copy of itself with a new value of $\alpha_{GF}(\mu)$ in a self-similar fashion.

The single most important question that must be answered before the problem of confinement can be settled is the behavior of the effective coupling  $\alpha_{GF}(\mu)$ in the limit $\mu \rightarrow 0$.

\section{Flow to confinement}

Let us consider the pure SU(3) gauge theory first. We do not expect any conceptual differences, as the effect of virtual gluons dominates over internal quark loops for a small number of flavors. In this case the gluon condensate is the only fundamental mass parameter of the theory. The gluon condensate we refer to is~\cite{Shifman:1978bx}
\begin{equation}
  G = \frac{\alpha_S}{\pi}\, \langle F_{\mu\nu}^a F_{\mu\nu}^a \rangle = 2\, \frac{\alpha_S}{\pi}\, \textrm{Tr} \,\langle F_{\mu\nu} F_{\mu\nu} \rangle 
  \label{G}
\end{equation}
($S$: scheme). We consider $G$ to be properly renormalized. When expressed in terms of gauge fields generated by the gradient flow, as we will consider, the gluon condensate does not require renormalization~\cite{Luscher:2011bx}. That has been utilized in~\cite{Artz:2019bpr}, where $\langle E(t)\rangle$ has been calculated to three loops for small flow times. This is to be contrasted with the condensate at flow time zero, which is highly divergent~\cite{Horsley:2012ra,Bali:2014sja}.

The gluon condensate $G$ anomalously breaks scale invariance, meaning the quantum theory does not inherit the scale symmetry of the classical action. It can be connected to the vacuum expectation value of the trace of the energy momentum tensor, the trace or scale anomaly,
\begin{equation}
  \langle \Theta_{\lambda\lambda}\rangle = \frac{\beta(\alpha_S)}{\alpha_S}\, \left\langle \frac{1}{4g_0^2}F_{\mu\nu}^a F_{\mu\nu}^a \right\rangle_R \,,
  \label{emt}
\end{equation}
which sometimes is taken for the gluon condensate~\cite{Banks:1981zf}. The subscript $R$ denotes renormalized fields defined in the scheme $S$. The trace (\ref{emt}) is obtained by taking the functional derivative of the action with respect to the metric tensor $g_{\mu\nu}$. In practice this means that the derivation is limited to zero flow time. The computation of (\ref{emt}) is highly nontrivial. In~\cite{Suzuki:2013gza} an attempt has been made to compute the renormalized energy momentum tensor from small flow times. %As we shall see, the expression (\ref{emt}) is definitely not scale invariant under the gradient flow. Beyond perturbation theory no result is known for finite flow time. %For a small flow time expansion see~\cite{Suzuki:2013gza}. 
%Thus, $G$ can be alternatively defined by a short flow-time expansion~\cite{Artz:2019bpr}.

We henceforth understand scale invariance to mean invariance under changes of $t = 1/8\mu^2$, with $g_0^2$ held fixed. The key point is that the renormalized gluon condensate $G$ is independent of the scale parameter $\mu$. This is known to be the case in perturbation theory~\cite{Kluberg-Stern:1974iel,Tarrach:1981bi}. A nonperturbative derivation was given in~\cite{Shifman:1978bx,Novikov:1979va,Graziani:1984cs,Suzuki:2018vfs,Beneke:2025hlg} utilizing the operator product expansion (OPE). In the Appendix I provide an alternative proof, in which $\langle G^a_{\mu\nu}G^a_{\mu\nu}\rangle$ is calculated directly from the functional integral at finite flow time. It leads to
\begin{equation}
  \langle G^a_{\mu\nu}G^a_{\mu\nu}\rangle \, \propto \, 1/t \,.
\end{equation}
Together with (\ref{coupling}) this yields $G \propto t^2 \langle  G_{\mu\nu}^a G_{\mu\nu}^a \rangle^2 = \textrm{constant}$. In addition, a loophole in the derivation of~\cite{Shifman:1978bx,Novikov:1979va} is closed, using the gradient flow as a source of regularization. The issue was also addressed in a recent publication~\cite{Beneke:2025hlg}.
%In the Appendix I present a proof that uses the gradient flow as a source of regularization.

This is put into practice by writing 
\begin{equation}
  G = \frac{\alpha_{GF}(\mu)}{\pi}\, \langle G_{\mu\nu}^a G_{\mu\nu}^a \rangle = 4 \frac{\alpha_{GF}(\mu)}{\pi}\, \langle E(t) \rangle 
  \label{GF}
\end{equation}
in the gradient flow scheme, with the result  
\begin{equation}
  \mu \,\frac{\partial\, G}{\partial\, \mu} = -2t \, \frac{\partial\, G}{\partial\, t} = 0 \,.
\end{equation}
The effect of the gradient flow is that as the time increases, higher frequencies are integrated out step by step, resulting in a decrease of energy density $\langle E(t) \rangle$, which is compensated for by an increase of $\alpha_{GF}(\mu)$. Using (\ref{coupling}), we can express the gluon condensate entirely in terms of the running coupling,
\begin{equation}
 G = \frac{3}{t^2} \, \left(\frac{\alpha_{GF}(\mu)}{\pi}\right)^2 = \frac{192}{\:\pi^2}\, \alpha_{GF}^2(\mu)\, \mu^4\,.
\end{equation}
Thus
\begin{equation}
  \mu \,\frac{\partial\, G}{\partial\, \mu} = \frac{384}{\:\pi^2}\, \alpha_{GF}(\mu)\, \Big(2 \alpha_{GF}(\mu) + \beta(\alpha_{GF})\Big)\; \mu^4 = 0\,.
\end{equation}
Altogether this leads to
\begin{equation}
  \beta(\alpha_{GF}) = - 2\, \alpha_{GF} \,, \quad \alpha_{GF}(\mu) = \frac{\Lambda_{GF}^2}{\mu^2} 
  \label{coupling2}
\end{equation}
with 
\begin{equation}
  \Lambda_{GF}^2 = \frac{\pi}{8} \sqrt{\frac{G}{3}} \,,
  \label{GL}
\end{equation}
where $\Lambda_{GF}$ is the lambda parameter in the gradient flow scheme. In~\cite{Schierholz:2024lge} I have shown how to connect (\ref{coupling2}) analytically to the perturbative result at high energies, $\alpha_{GF}(\mu) \simeq 1/4\pi b_0 \log\,(\mu^2/\Lambda_{GF}^2)$. It follows that $\Lambda_{GF}$ in (\ref{coupling2}) is identical to the lambda parameter in the perturbative equations. 
The infrared behavior of the running coupling is universal. Under a change of renormalization scheme, $GF \rightarrow S$, we have~\cite{Schierholz:2024lge}  
\begin{equation}
  \log {\frac{\Lambda_S}{\Lambda_{GF}}} = - \int_{\alpha_{GF}(\mu)}^{\alpha_{S}(\mu)} \, \frac{d\alpha}{\beta(\alpha)} \,,
  \label{scheme}
\end{equation}
where the measure $d\alpha/\beta$ is scheme invariant. Knowing $\alpha_{GF}(\mu)$ and the $\beta$ function for small values of $\mu$, we can solve (\ref{scheme}) for $\alpha_S(\mu)$. The result is
\begin{equation}
  \alpha_S(\mu) = \frac{\Lambda_S^2}{\mu^2} \,.
  \label{S}
\end{equation}
Of particular phenomenological interest is the $\MSbar$ scheme. In the pure gauge theory $\Lambda_{\MSbar} = 0.534\, \Lambda_{GF}$. The lambda parameter now is not just an integration constant, but has a tangible meaning. This is a significant result, which has been suggested before~\cite{Hill:2005wg}, but has been quantified here for the first time.

%Here we have made use of the fact that (\ref{coupling2}) is analytically connected to the perturbative result at high energies with $\alpha_{GF}(\mu) \simeq 1/4\pi b_0 \log\,(\mu^2/\Lambda_{GF}^2)$~\cite{Schierholz:2024lge}. %y by analyticity expressionsIn~\cite{} An analytical connection between (\ref{coupling2}), valid at low energies, and the corresponding two-loop result at perturbative scales was presented in~\cite{} and will not be repeated here.
%scheme with
%\begin{equation}
%  \Lambda_{GF} = 1.873 \, \Lambda_{\MSbar} \,.
%\end{equation}

The result (\ref{S}) that $\alpha_S(\mu)$ increases linearly with the inverse power of $\mu^2$ beyond any bound, bypassing the Landau pole, was long awaited. In the literature it is referred to as infrared slavery. This is clear evidence of color confinement. In order to prove confinement beyond doubt, it would be sufficient to prove the existence of a nonvanishing gluon condensate $G$. It is known for a long time that the perturbative vacuum is not the true ground state of the theory. The vacuum with zero field strength $\langle F_{\mu\nu}^aF_{\mu\nu}^a\rangle$, for example, was found to be unstable~\cite{Savvidy:1977as}. This is evident in the fact that the Euclidean effective action is minimized by a nontrivial condensate.

This is a remarkable turn. The QCD vacuum has many facets, depending on the scale parameter $\mu$. A homogeneous vacuum would be unstable as well~\cite{Nielsen:1978rm}, indicating that the vacuum has granular structure. This has been brought to light in numerous work, emphasizing the role of monopoles and vortices. While the authors argue about which of them is ultimately responsible for confinement, it appears that the microscopic properties of the vacuum are largely averaged out when seen from large distances, leaving many possibilities behind at intermediate scales. The solution is characterized by a single, scale-invariant limit under the renormalization group flow, the infrared fixed point $1/\alpha_S = 0$. This is known as universality. Theories in a given universality class could look very different microscopically, but will end up looking the same at long distances.
%\vspace*{-0.25cm}

\section{Numerical verification}

Many of these results have already been documented in previous publications~\cite{Nakamura:2021meh,Schierholz:2024lge}. The computations are based on the Wilson gauge field action at $\beta = 6.0$, corresponding to $g_0^2 = 1$ and lattice spacing $a = 0.082\, \textrm{fm}$. Three hypercubic volumes $V = L^4$ were considered, $L/a =16, 24$ and $32$. The clover definition of $G_{\mu\nu}$ was used. In the following I will reformulate some of the results to illustrate and substantiate the claims made in the previous section.

\begin{figure}[b!] \vspace*{-0.5cm}
  \begin{center}
  \epsfig{file=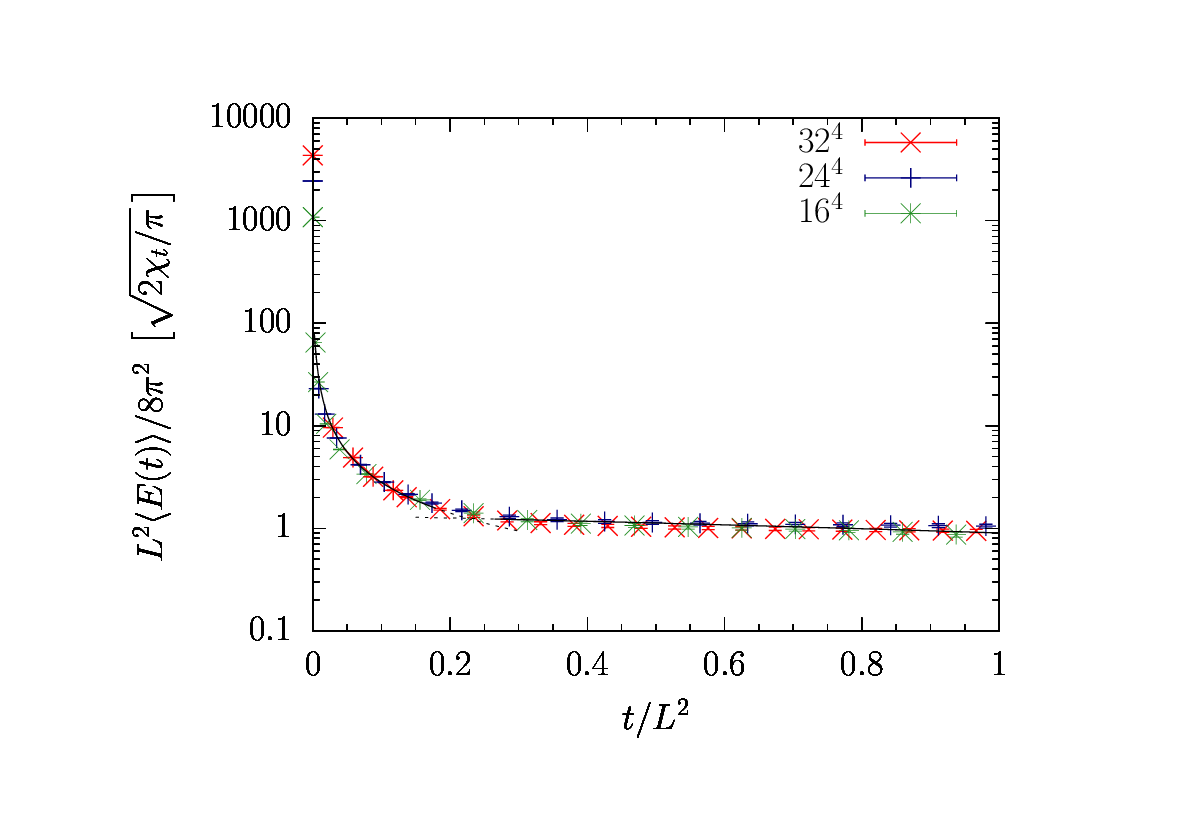,width=12.85cm,clip=}
  \end{center} \vspace*{-1.25cm}
  \caption{The energy density $L^2 E(t)/8\pi^2$ in units of $\sqrt{2\chi_t/\pi}$ as a function of $t/L^2$, where $\chi_t = \langle Q^2 \rangle/V$ has been taken from~\cite{Nakamura:2021meh}. The curves are a fit of the form $c/t$ to the left-hand data ($t/L^2 \lesssim 0.2$), and a linear fit to the right-hand data.}
  \label{fig1}
\end{figure}

Let us start with the energy density $\langle E(t)\rangle$. In Fig.~\ref{fig1} I plot $L^2 \langle E(t)\rangle/8\pi^2$ against $t/L^2$ for our three volumes, where $L^2 E(t)/8\pi^2$ has been expressed in units of $\sqrt{\,2\,\chi_t/\pi}$ for reasons that will become clear in a moment. The immediate observation is that the data for the different volumes fall on a single line. This tells us two things. First, $\langle E(t)\rangle \propto 1/t$ for $t/L^2 \lesssim 0.2$, in accord with $\alpha_{GF} \simeq 8 \Lambda_{GF}^2 t$. At $t/L^2 \approx 0.21$, corresponding to $\sqrt{8 t} \approx 1.25\, L$, the averaging process reached its limit, resulting in an infrared cut-off on the renormalization group transformation. (Note that on the $32^4$ lattice this is far above the flow times utilized in~\cite{Nakamura:2021meh}.)
%group ..The ... process appears to be exhausted at What is new is that this relationship holds up to $\sqrt{t} \approx 0.5\,L$, which represents an infrared cut-off on renormalization group transformations. . This means, for example, that on a $5\, \textrm{fm}$ lattice the lowest accessible scale parameter is about $\mu \approx 30\,\textrm{MeV}$.
Secondly, $\langle E(t)\rangle$ does not decrease anymore for $t/L^2 \gtrsim 0.2$, but levels off to an approximately constant value, 
\begin{equation}
  V \,E(t) = 8\, \pi^2 \sqrt{\,\frac{2}{\pi}\, \langle Q^{2} \rangle} \approx  8\, \pi^2 \, \langle\, |Q|\,\rangle \,,
  \label{inst}
\end{equation}
the average classical action. The occurrence of mixed states of instantons and anti-instantons has been ruled out here~\cite{Schierholz:2024var}. The linear curve in Fig.~\ref{fig1} is not exactly horizontal, because we lost some charges in the process, especially in the sectors of large $Q$, probably due to dislocations~\cite{Gockeler:1989qg}. Apart from that, (\ref{inst}) is what one expects for a set of noninteracting (anti-)instantons with topological charge $|Q|$, given the action for a single (anti-)instanton  $V \,\langle E\rangle = 8 \pi^2$ and the average charge and normalization factor
\begin{equation}
  \langle |Q|\rangle = \frac{1}{\sqrt{2 \pi\,\langle Q^2\rangle}} \int dQ\, |Q|\, \exp\left\{-Q^2/2\langle Q^2\rangle\right\} = \sqrt{\,\frac{2}{\pi}\, \langle\, Q^{\,2}\, \rangle} \,.
  \label{gauss}
\end{equation}
The topological charge, which counts the number of (anti-)instantons, is modeled as a random variable, whose probability density is described by a Gaussian distribution. This corresponds to the distribution of topological charge expected and observed in the large volume~\cite{Schierholz:2024var,Durr:2025qtq}. See also the Appendix for more information on the Gaussian distribution. On the plateau, $t/L^2 \gtrsim 0.2$, the energy density $\langle E(t)\rangle$ vanishes with the inverse root of the volume, while the topological susceptibility $\chi_t$ stays constant, as a result of $\partial\, Q/\partial \, t = 0$. %See the Appendix.

The picture is not expected to change quantitatively in the continuum limit, except for possible $O(a^2)$ effects. The `constant' line nicely demonstrates the existence of (anti-)instantons. This is not new~\cite{Ilgenfritz:1985dz}, but the reference to a physical scale and the onset of classical behavior is. In~\cite{Nakamura:2021meh} the energy density has been broken down according to the topological charge $|Q|$. %The flow time $t$ has no obvious physical meaning beyond $t \approx 0.2 \, L^2$. 

\begin{figure}[b!] \vspace*{-0.5cm}
  \begin{center}
  \epsfig{file=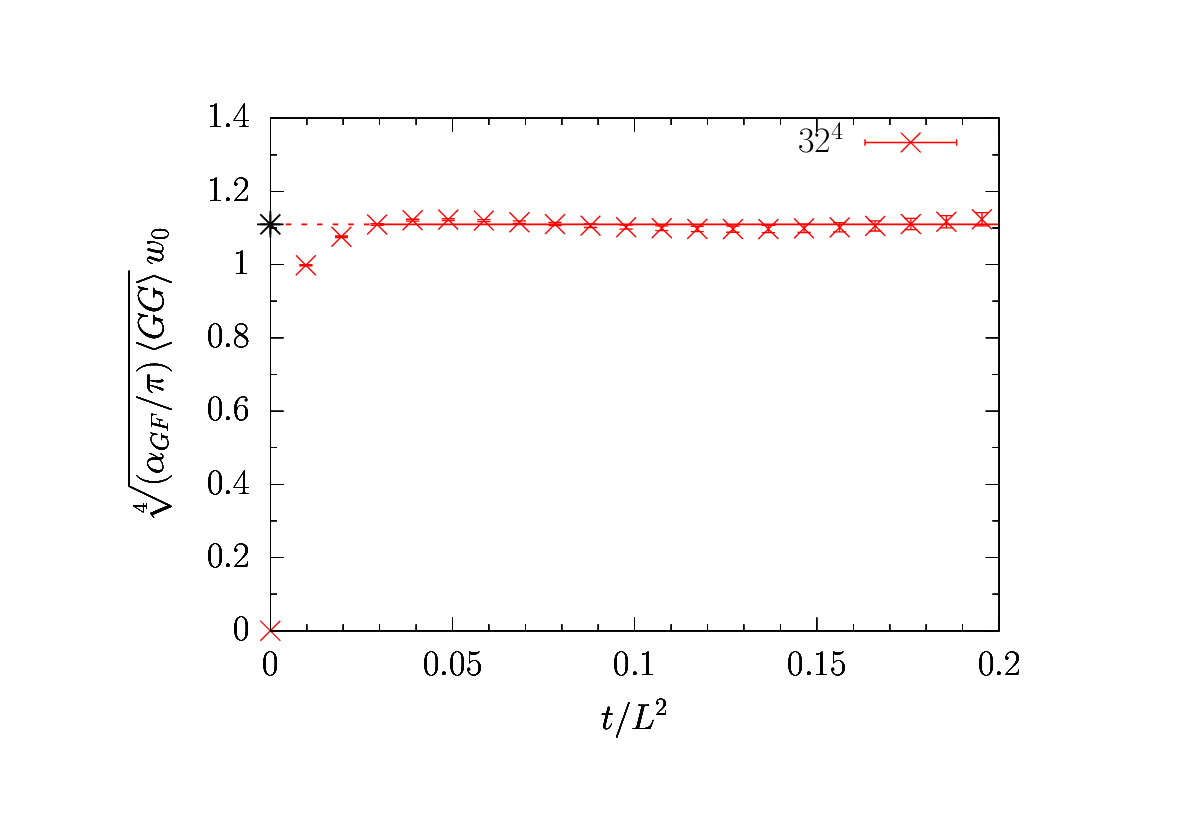,width=12cm,clip=}
  \end{center} \vspace*{-1.25cm}
  \caption{The gluon condensate on the $32^4$ lattice in units of the flow parameter $w_0$ as a function of the dimensionless quantity $t/L^2$ in the gradient flow scheme.}
  \label{fig2}
\end{figure}

We now turn to the gluon condensate, on which our derivations are essentially based. On the lattice, it suffers from perturbative mixing with the unit operator, which results in an ultraviolet divergence. Recently, the perturbative contribution has been computed to $20$~\cite{Horsley:2012ra} and $32$ loops~\cite{Bali:2014sja} in numerical stochastic perturbation theory and subtracted from the Monte Carlo data to arrive at a finite answer. In Fig.~\ref{fig2} I show the gluon condensate (\ref{GF}) as a function of flow time on the $32^4$ lattice. Remember, sensible flow times are $t/L^2 \lesssim 0.2$, which corresponds to $t/a^2 \lesssim 200$ on this lattice. No attempt has been made to correct for lattice artifacts. As was expected, the gluon condensate is constant within the error bars for `nonperturbative' values of the flow time. A minor deviation in the transition region at $t/L^2 \approx 0.2$ should not surprise us. At small flow time the energy density $\langle E(t)\rangle$ suffers from perturbative contributions. `Nonperturbatve' means once the perturbative contributions have died out and $\alpha_{GF}(\mu)$ has assumed its asymptotic form, which happens to be at $t/L^2 \gtrsim 0.02$. This matches the short flow-time behavior reported in~\cite{Schierholz:2024lge}. The lattice renormalized gluon condensate is obtained by the continuation of the straight line to $t=0$, marked by the black star.

In the $\MSbar$ scheme the gluon condensate becomes
\begin{equation}
  G = \frac{192}{\pi^2} \, \Lambda_{\MSbar}^4 = 19.45 \, \Lambda_{\MSbar}^4\,.
  \label{GMSbar}
\end{equation}
This is consistent with the extrapolated number in  Fig.~\ref{fig2}, given that $w_0 \Lambda_{\MSbar} = 0.212$~\cite{Schierholz:2024lge}. For $\Lambda_{\MSbar} = 0.255(4) \, \textrm{GeV}$~\cite{FlavourLatticeAveragingGroupFLAG:2024oxs} this gives $G = 0.082(3) \, \textrm{GeV}^4 \equiv (535(4)\,\textrm{MeV})^4$. Recent lattice calculations give a mixed answer. In~\cite{Horsley:2012ra} the result $G = 0.028(3)\, \textrm{GeV}^4$ was reported, while the authors of~\cite{Bali:2014sja} quote $-(2\beta(\alpha)/\beta_0 \, \alpha) \, \langle F_{\mu\nu}^a F_{\mu\nu}^a\rangle = (1+ O(\alpha))\, G = 24(8) \,\Lambda_{\MSbar}^4$, which is in agreement with our result (\ref{GMSbar}). This brings us close to a reliable figure. Both numbers refer to the pure gauge theory. A first, phenomenological estimate from sum rules gave $G \approx 0.012(3) \, \textrm{GeV}^4$~\cite{Shifman:1978bx}, which is subject to considerable theoretical uncertainty however.

\begin{figure}[b!] %\vspace*{-0.75cm}
  \begin{center}
  \epsfig{file=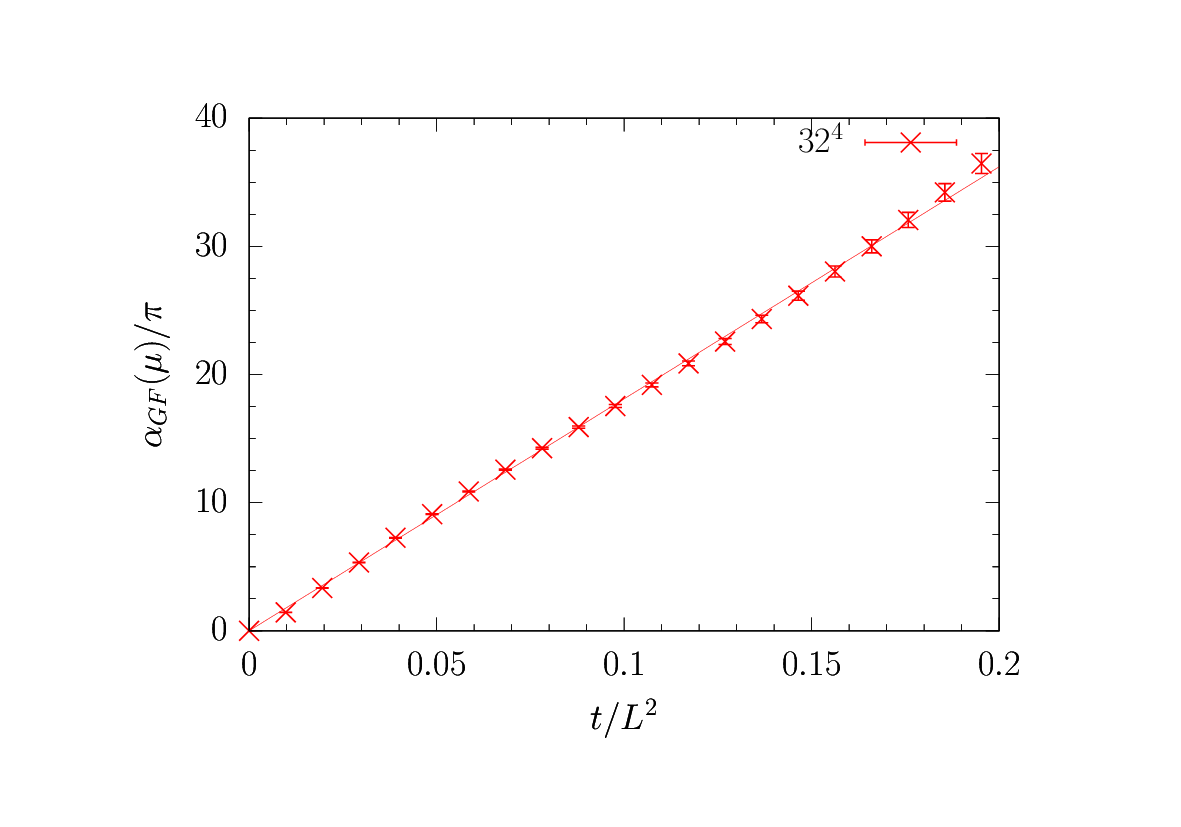,width=11.8cm,clip=}
  \end{center} \vspace*{-1.25cm}
  \caption{The running coupling $\alpha_{GF}(\mu)$ on the $32^4$ lattice as a function of $t/L^2$. On this lattice $\mu \approx 60 \, \textrm{MeV}$ at the border value.}
  \label{fig3}
\end{figure}

To finish off, in Fig.~\ref{fig3} I show the running coupling constant $\alpha_{GF}(\mu)$ as a function of $t/L^2$ on the $32^4$ lattice. We have already seen in Fig.~\ref{fig1} that $\alpha_{GF}(\mu)$ is independent of the volume. Apart from a few wobbles close to the infrared cut-off, the result is a remarkably straight line. The slope of the line is $\sqrt{G/3}\, L^2$, which is consistent with the result for $G$ given in Fig.~\ref{fig2}. This confirms our analytical result (\ref{coupling2}). The flow equations suggest that the gauge fields are averaged over a range of $\sqrt{8 t}$. In rectangular geometry that would limit $t$ to $t/L^2 \lesssim 1/8$. However, if rotation invariance is restored, the limit might be a factor two bigger.

\section{Corollary}

Our derivations thus far represent a significant departure from previous approaches to confinement. Yet many of the characteristics of confinement can be traced back to a nonvanishing gluon condensate and its polarizing effect on the vacuum, without detailed understanding of the underlying mechanism. In the following I give a few examples.

%are conditioned
%\subsection*{Topology}

\begin{figure}[b!] \vspace*{-0.75cm}
  \begin{center}
  \epsfig{file=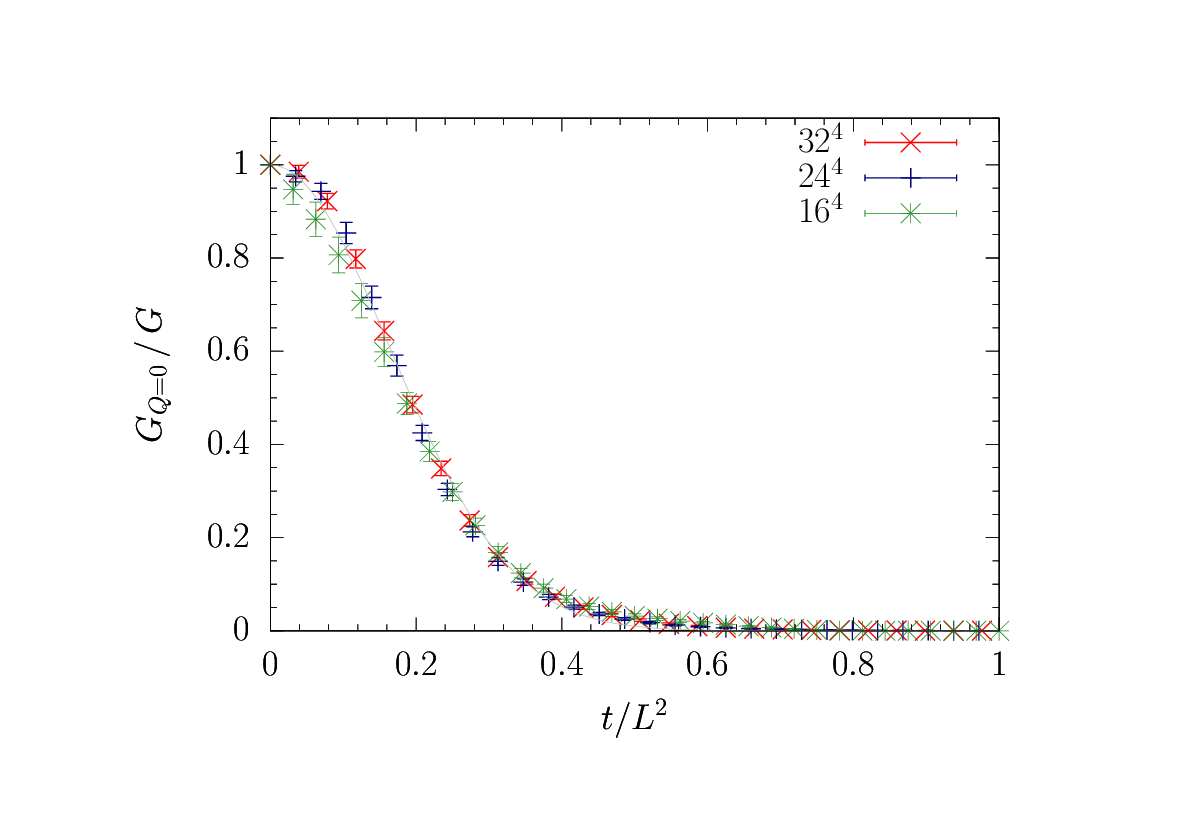,width=12cm,clip=}
  \end{center} \vspace*{-1.25cm}
  \caption{The ratio of the gluon condensate $G_{Q=0}$ for topological charge zero to the full gluon condensate $G$ as a function of $t/L^2$. The solid curve is a Gaussian fit.}
  \label{fig4}
\end{figure}

As was mentioned already, the homogeneous vacuum is unstable. A quantity that measures the granularity and fluctuations of the vacuum is the topological susceptibility $\chi_t$, which plays a role in various aspects of confinement and strong CP violation. Furthermore, it acts as an auxiliary source of mass generation. The two condensates are connected, as can be seen in Fig.~\ref{fig1}. At $t/L^2 \approx 0.21$, which marks the intersection point between renormalization group decimation and constant behavior, indicated by the dashed lines, the energy density $\langle E(t)\rangle$ can be identified with the minimal classical action (\ref{inst}). From this follows
\begin{equation}
  L^2 \, \langle E(t) \rangle \, \frac{t}{L^2} \approx 0.21 \times 8 \pi^2 \, \sqrt{\frac{2}{\pi} \, \chi_t} \,.
\end{equation}
Substituting $\sqrt{3 G/16}$ for $\langle E(t)\rangle t$ finally gives the result for $G$ in the $\MSbar$ scheme 
\begin{equation}
  \sqrt{G} \approx 0.21 \times 41.5 \, \sqrt{\chi_t} = 8.7 \, \sqrt{\chi_t} \;\; \therefore \;\; G = 75.7 \, \chi_t \,.
  \label{Gchi}
\end{equation}
Given that $\chi_t = (180(4) \, \textrm{MeV})^4$~\cite{Ce:2015qha}, and our result (\ref{GMSbar}) for $G$, the expected ratio is $G/\chi_t = (535/180)^4=78$, which makes (\ref{Gchi}) a perfect match. This is to be compared with results from the literature, $G/\chi_t = 22$~\cite{Novikov:1981xi} and $G/\chi_t = 49.5$~\cite{Halperin:1997vj}. Analytically, $t/L^2 = 0.030\, \Lambda_{GF}^2/\sqrt{\chi_t}$ at the intersection point, assuming a horizontal line in Fig.~\ref{fig1}. An interesting question is how much instantons contribute to the gluon condensate. In Fig.~\ref{fig4} I show the condensate for topological charge zero divided by the total value, $G_{Q=0}/G$. At $t/L^2 = 0.21$ we find $G_{Q=0}/G \approx 0.4$,  which suggests that approximately $60\%$ of the magnitude of the gluon condensate is due to instantons. This is not surprising, as instantons have a finite action and may outweigh the low-frequency gluonic modes, depending on the volume and the scale parameter $\mu$.

Interestingly, the topologically trivial condensate $G_{Q=0}$ is not invariant under renormalization group transformations in the finite volume. The ratio of $G_{Q=0}$ to $G$ is well approximated by a Gaussian distribution,
\begin{equation}
  \frac{G_{Q=0}}{G} \approx \exp\{-c\, t^2/L^4\}\,,
\end{equation}
with $c \approx 18.5$, shown by the solid curve in Fig.~\ref{fig4}. Minor finite size corrections seen on the $16^4$ lattice should not disturb us. Following the derivations in the previous section, we arrive at the running coupling in the trivial sector 
\begin{equation}
  \alpha_{GF}^{Q=0}(\mu) \approx  \alpha_{GF}(\mu) \, \exp\{-(c/2)\, t^2/L^4\}\,.
\end{equation}
This tells us that if the path integral is restricted to configurations with trivial topology, $Q=0$, one obtains a different `theory' involving long-range correlations. The long-range correlations occur if $\mu L$ is small. However, if $\mu$ takes on any nonzero value, the correlations disappear as the volume tends towards infinity. This is in broad agreement with the observation that the density of zero modes vanishes with the inverse square root of the volume~\cite{Schierholz:2024var}. %We conclude that the `theory' cannot be formulated sensibly in local terms. %Whether the `theory' confines or not is an academic question. It has no anomaly and thus is not a viable theory. %Care must be taken when assessing the long-distance properties under these circumstances.  

%If, however, $\mu$ is kept fixed while the volume is taken to infinity it appears to be irrelevant whether the path integral is one restthe path integral is restricted to ... The That emphasizes the importance of vacuum topology for is is equivalent to 

%This tells us that the gluon condensate is entirely saturated by the topological susceptibility. From the renormalization group point of view this is not really surprising. A heuristic derivation~\cite{Halperin:1997vj}, referring to a different regularization scheme, gave the number $\sqrt{G} = 7.0 \, \sqrt{\chi_t}$. 

%A nonvanishing string tension $\sigma$ is widely considered a criterion for confinement. In essence, it is a concrete way of describing the confinement phenomenon.

A nonvanishing string tension $\sigma$ is considered a stringent requirement for confinement. The renormalization group provides a framework for calculating the string tension diagrammatically. Infrared singularities that arise in scattering amplitudes can be collected by a reorganization of perturbation theory into insertions of single gluons between external charges, with the respective effective couplings $\alpha_S(q)$ taken at the gluons' momentum $q$, in a fashion that resembles the low-order results~\cite{Crewther:1978qs}. At small virtualities we expect only tree-level diagrams to contribute. In the infrared, the static potential can thus be approximated by 
\begin{equation}
  V(q)= -\frac{16 \pi}{3}\,\frac{\alpha_V(q)}{q^2} \,,
  \label{potq}
\end{equation}
where $\alpha_V(q)$ is the effective coupling in the $V$ scheme, $\alpha_V(q) = \Lambda_V^2/q^2$. This can be envisaged as the exchange of a single dressed gluon, forming a string.
%When it comes to phenomenological applications, (\ref{potq}) should be considered the lower limit of the static potential. The conversion factor to the $\MSbar$ scheme is $\Lambda_V = 1.61 \, \Lambda_{\MSbar}$~\cite{Schroder:1998vy}.
In the infinite volume the gluon momentum $q^2$ is not bounded from below, and thus can be integrated to zero. The Fourier transformation of $V(q)$ to configuration space then gives~\cite{Jezabek:1998wk}
\begin{equation}
  V(r) = -\frac{1}{(2\pi)^3} \int d^3\mathbf{q} \; e^{i\,\mathbf{q r}} \; \frac{16\pi}{3}\, \frac{\alpha_V(q)}{\mathbf{q}^2}\; \underset{r\, \gg\, 1/\Lambda_V}{=} \,\frac{2}{3}\, \Lambda_V^2 \,r\;\; \therefore \;\; \sigma = \frac{2}{3}\, \Lambda_V^2 \,.
  \label{potr}
\end{equation}
The charmonium and bottomonium spectrum is very well described by $\Lambda_V \approx 0.47\, \textrm{GeV}$~\cite{Richardson:1978bt}, which is to be compared with~\cite{Schroder:1998vy} $\Lambda_V = 1.61\, \Lambda_{\MSbar} = 0.41(1)\, \textrm{GeV}$ (assuming that $\alpha_V(q)$ is nonperturbatively defined~\cite{Necco:2001gh}). 

Confinement is often associated with the existence of a mass gap. But confinement and the existence of a mass gap is not the same. Having a gap in the spectrum does, for example, not exclude massive free gluons and quarks. Furthermore, the existence of a massless Goldstone boson would contradict the mass gap criterion. Leaving this possibility aside, a mass gap appears to be a necessary condition for confinement, but it is not sufficient on its own. It has been suggested~\cite{Mack:1977xu} that the gluon condensate gives rise to a set of Higgs scalars, which in turn give mass to the gauge bosons, the mass gap. However, the Higgs scalars cannot screen the color charge of quarks. That was the missing point in~\cite{Mack:1977xu}, and the reason why the model was ultimately discarded. This shortcoming has now been overcome, so that all fields in the fundamental representation -- quarks, tetraquarks and the like -- are subject to the strong force $\alpha_S(\mu) \underset{\mu \rightarrow 0}{\rightarrow} \infty$. 

The essence of the Higgs mechanism is to derive a local Lagrangian containing only color singlet gauge fields. Higgs scalars that offer themselves are~\cite{Mack:1977xu}
\begin{equation}
  \begin{tabular}{cc}
  $\begin{split}
%  $\begin{align}
  \Phi_1(x) &= F_{\mu\nu}(x)\, F_{\mu\nu}(x) - \textrm{Tr}\,, \\  \Phi_2(x) &= F_{\mu\nu}(x)\, F_{\nu\rho}(x)\, F_{\rho\mu}(x)- \textrm{Tr} \,,
  \end{split}$ &  
%  \end{align}$ &  
  $\begin{split}
  \Phi_1^a(x) &= \mathrm{Tr}\, \lambda_a \,F_{\mu\nu}(x)\, F_{\mu\nu}(x) \,,\\  \Phi_2^a(x) &= \mathrm{Tr}\, \lambda_a \,F_{\mu\nu}(x)\, F_{\nu\rho}(x)\, F_{\rho\mu}(x) \,.
  \end{split}$
  \end{tabular}
  \label{hs}
\end{equation}
The transformation to gauge invariant field variables requires to define a unique gauge transformation $g(x) \in \textrm{SU(3)}/\textrm{Z}_3$. This is a two step process. We first diagonalize $\Phi_1(x)$,
\begin{equation}
  g(x)\, \Phi_1(x)\, g(x)^{-1} = \hat{\!\Phi}_1(x) = \textrm{diag}\left(\phi_1^1,\phi_1^2,\phi_1^3\right)\,,
  \label{abel}
\end{equation}
which determines $g(x)$ up to left multiplication by
\begin{equation}
  l(x) = \textrm{diag}\left(\exp{(i\alpha_1)},\exp{(i\alpha_2)},\exp{(i\alpha_3)}\right)\,, \quad \sum_i \alpha_i = 0 
\end{equation}
with $l(x) \in \textrm{U(1)}\times\textrm{U(1)}$, the Cartan or largest abelian subgroup of SU(3). The transformation $g(x)$ is uniquely fixed by considering \begin{equation}
  g(x)\, \Phi_2(x)\, g(x)^{-1} = \hat{\!\Phi}_2(x)
\end{equation}
and demanding that the elements of $\,\hat{\!\Phi}_2$ on the second diagonal, $(\,\hat{\!\Phi}_2)_{12}(x)$ and $(\,\hat{\!\Phi}_2)_{23}(x)$, are real and positive. See also~\cite{deForcrand:2000pg,Reinhardt:2000db}.
We need to have $\Phi_1(x), \,\Phi_2(x) \neq 0$ only where $F_{\mu\nu}(x) \neq 0$. It is therefore possible that gluons can screen their color and acquire mass through a local Higgs mechanism. We then may define the gauge invariant quantities
\begin{equation}
  \begin{split}
 \hat{A}_\mu(x) &= g(x)\, A_\mu(x)\, g^{-1}(x) +  g(x)\, \partial_\mu\, g^{-1}(x) \,, \\%[0.5em]
 \hat{F}_{\mu\nu}(x) &= \partial_\mu \hat{A}_\nu(x) -  \partial_\nu \hat{A}_\mu(x) +\,\left[\hat{A}_\mu(x),\hat{A}_\nu(x)\right]
  \end{split}
\end{equation}
with the gauge field action \vspace*{-0.15cm}
\begin{equation}
  S_G = \frac{1}{2 g_0^2} \int d^4x\, \mathrm{Tr}\,\hat{F}_{\mu\nu} \hat{F}_{\mu\nu} \,.
\end{equation}
The final result is that the gauge fields $A_\mu$ combine with the Higgs scalars $\Phi_1,\,\Phi_2$ to form colorless massive vector particles. Furthermore, the Higgs scalars (\ref{hs}) shield the color electric field that binds quark and antiquark together, so as to form a tube-like structure, the string.%, rather than spreading out like in QED.

The transformation (\ref{abel}) fixes the nonabelian part of the gauge completely, which leaves us with abelian gauge fields, gluons, monopoles and quarks, forming the dual superconductor model of confinement~\cite{tHooft:1981bkw,Kronfeld:1987vd}. From a physical point of view, given the key role of the gluon condensate, I would suggest using the gauge (\ref{abel}) fixed by diagonalizing the Higgs scalar $\Phi_1$ rather than the maximally abelian gauge suggested in~\cite{Kronfeld:1987ri}. Both can be expected to give similar results, which a comparison of~\cite{Kronfeld:1987vd} and ~\cite{Kronfeld:1987ri} suggests.

Although this approach is conceptually appealing, it can hardly be developed into a proof of the mass gap. In the Appendix, I derive the mass gap under the assumption of a nonvanishing and finite gluon condensate from the gradient flow. Calling the lowest-lying state above the vacuum $\sigma$, the result is
\begin{equation}
  m_\sigma^2 = \frac{\pi^3}{576} \,\frac{\left\langle 0\big|\,\left(\alpha_{GF}/\pi\right)\, G_{\mu\nu}^a G_{\mu\nu}^a\,\big|\sigma\right\rangle^2}{\Lambda_{GF}^6}
   \,.
\end{equation}
The case $m_\sigma = 0$ can be excluded.

\begin{figure}[b!] \vspace*{-0.75cm}
  \begin{center}
  \epsfig{file=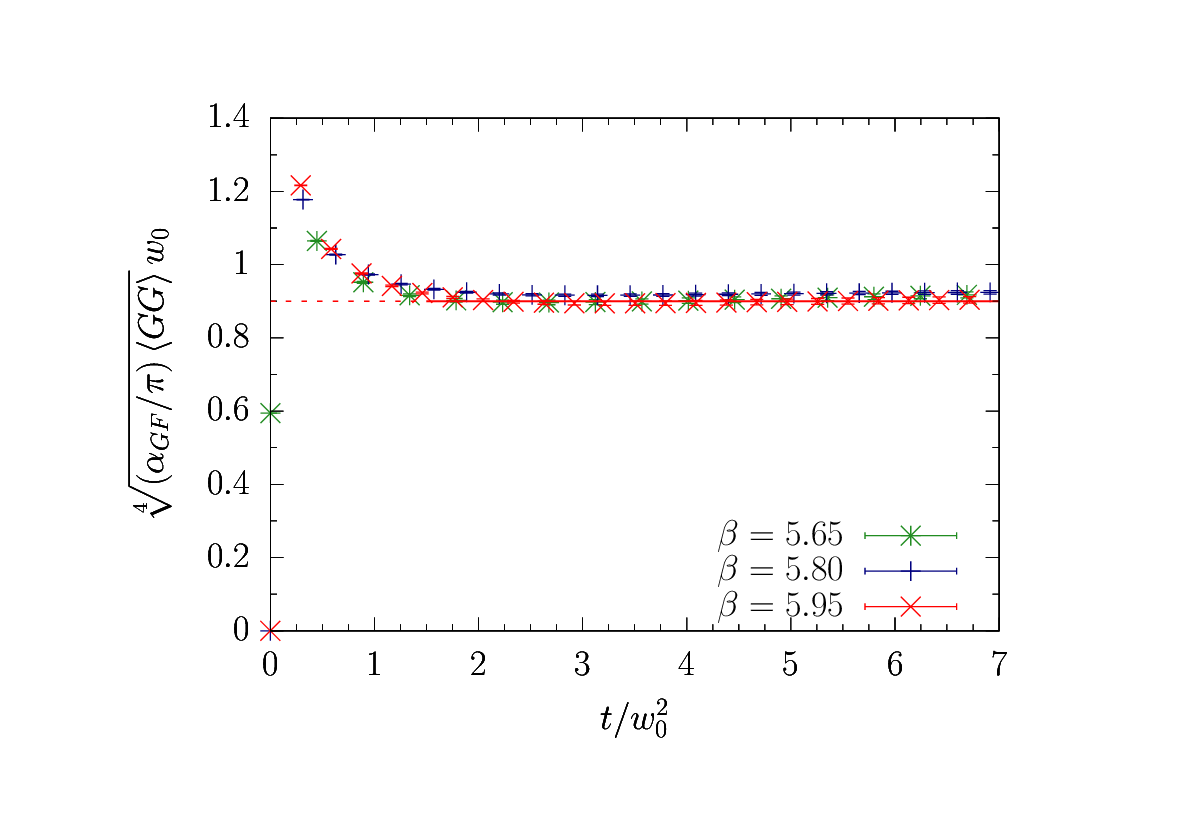,width=12cm,clip=}
  \end{center} \vspace*{-1.25cm}
  \caption{The gluon condensate in units of $w_0$ as a function of $t/w_0^2$ for three values of $\beta$ on the $32^3 \times 64$ ($\beta=5.65$) and $48^3 \times 96$ lattice ($\beta=5.8$ and $5.95$).}
  \label{fig5}
\end{figure}

%Any excitation of the vacuum is typically associated with the existence of particles. The mass gap is the smallest amount of energy needed to excite the vacuum from its ground state to a state where particles are present. 

%\begin{figure}[b!] \vspace*{-0.75cm}
%  \begin{center}
%  \hspace*{0.4cm}\epsfig{file=alphaSYM.ps,width=11.75cm,clip=}
%  \end{center} \vspace*{-1.25cm}
%  \caption{The running coupling $\alpha_{GF}(\mu)$ as a function of $t/w_0^2$ %for our three values of $\beta$.}
%  \label{fig6}
%\end{figure}

\section{Light flavors}

\begin{figure}[t!] \vspace*{-0.75cm}
  \begin{center}
  \hspace*{0.4cm}\epsfig{file=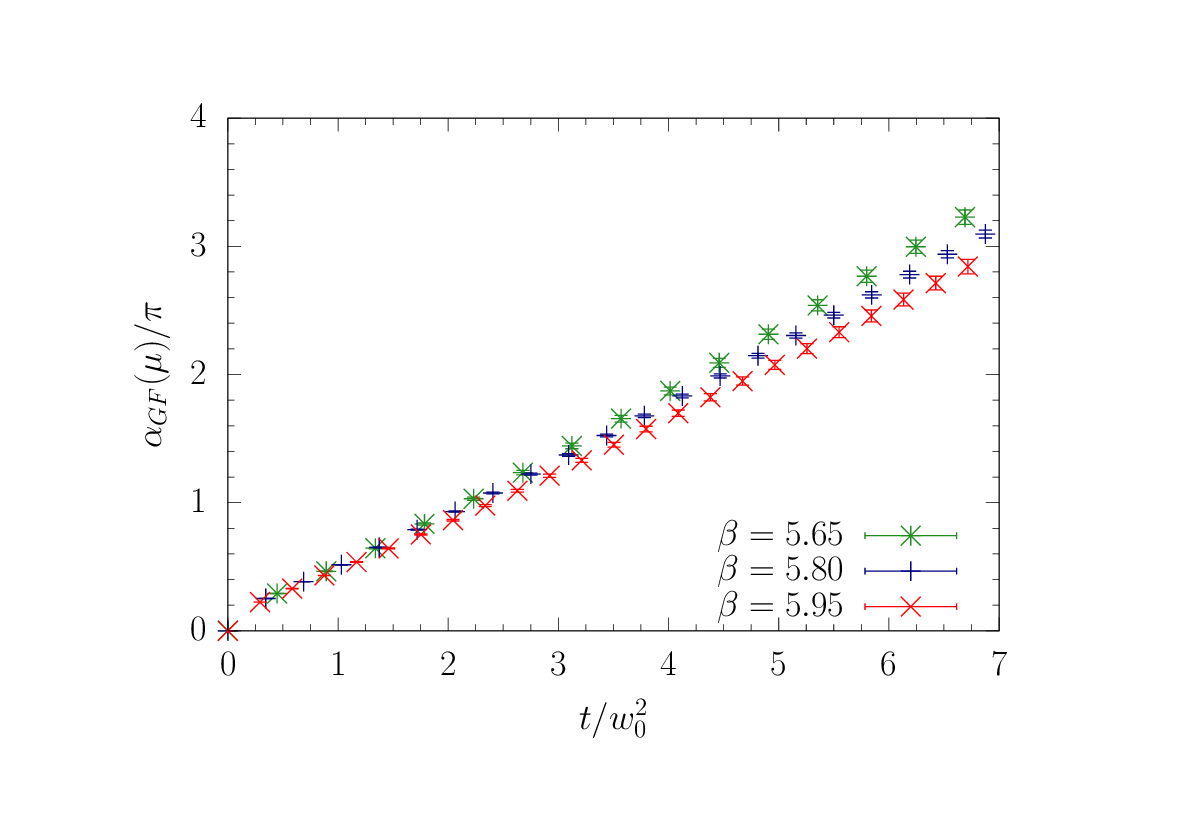,width=11.75cm,clip=}
  \end{center} \vspace*{-1.25cm}
  \caption{The running coupling $\alpha_{GF}(\mu)$ as a function of $t/w_0^2$ for our three values of $\beta$.}
  \label{fig6}
\end{figure}

We will now demonstrate that the same results essentially hold in the presence of a small number of dynamical quarks. However, there are a few factors that complicate the quantitative analysis exposed in~\cite{Zanotti:2021hma}. In previous work we generated flowed configurations with $2 + 1$ quark flavors, in an attempt to compute the gradient flow scales $\sqrt{t_0}$ and $w_0$~\cite{Bornyakov:2015eaa}, to which I refer here. The calculations were done at the SU(3) flavor symmetric point~\cite{Bietenholz:2011qq}, with the bare parameters $\beta$ and $\kappa$ held fixed. The calculations stopped at $t/a^2 = 80$. The first quantity to consider is the gluon condensate.
In Fig.~\ref{fig5} I show the gluon condensate $G$ as a function of $t/w_0^2$. The results are expressed in units of $w_0$ in order to combine the different $\beta$ values. We see that $G$ is independent of $t$ and about half the size of its quenched counterpart. The latter does not surprise. It goes hand in hand with a significant reduction in topological susceptibility $\chi_t$. %Remember that $\chi_t$ is independent of the flow time. 
Furthermore, in Fig.~\ref{fig6} I show the running coupling. Within the error bars, the running coupling is found to be a linear function of $t$ on the $48^3 \times 96$ lattice at $\beta = 5.95$, while the results on the smaller lattice at $\beta = 5.65$ show some finite size effects. We may conclude that infrared slavery persists in the presence of dynamical quarks. This should remain the case for as long as the gluon self-interaction outweighs the screening effect induced by the quarks.
It should be noted, however, that in order to keep the physics constant, such as the ratio of pion to nucleon mass, the bare quark masses must be adjusted as the flow time is increased~\cite{Zanotti:2021hma}. This also requires an adjustment of the bare coupling, which makes quantitative estimates, such as~\cite{Schierholz:2024lge}, difficult. We will return to this problem in a separate publication.

In~\cite{deTeramond:2024ikl} it has been argued that $\alpha_S(q)$ becomes near constant at small virtualities, suggesting approximate conformal behavior in the infrared. This contradicts our results, assuming that $\alpha_S(q)$ is correctly reproduced from the experimental data. Extracting $\alpha_S(q)$ for small $q^2$ from the Bjorken and Gross--Llewellyn-Smith sum rules, for example, is difficult because the leading-twist Wilson coefficient mixes with the matrix elements of higher-twist operators, making $\alpha_S(q)$ practically unobservable beyond the perturbative domain. This is an unsolved problem with as yet unknown consequences for $\alpha_S(q)$.  

%The quantitative assessment By comparing Figs.~\ref{fig6} and \ref{fig3} we find that $\alpha_{GF}(\mu)$ is approximately $40\%$ smaller than the quenched result. Note, however, that this refers to an unspecified value of the quark mass. It is not clear to me at present how to translate the result into a value of $\Lambda_{\MSbar}$, which is not a valid parameter beyond perturbation theory in full QCD. What we can say is that infrared slavery persists in the presence dynamical quarks. This should remain the case for as long as the gluon self-interaction outweighs the screening effect induced by the quarks.%The important point to make we tentatively conclude

% all have been named criterion for confinement
%\section{Infrared slavery}

%\section{Mass gap}

\section{Conclusions}

In this work we have identified the condition for confinement, and presented a first principles approach to determining the long-distance properties of QCD, thereby establishing robust measures for verification. A sufficient condition for confinement appears to be the existence of a nonvanishing gluon condensate $(\alpha_S/\pi)\, \langle G_{\mu\nu}^a G_{\mu\nu}^a\rangle$. It is associated with an `outer' mass scale that is essential for the existence of QCD. While the existence of a gluon condensate is not directly proven, its presence is strongly suggested by observations that cannot be explained without it, such as the anomalous breaking of scale invariance. It is known though that the perturbative vacuum is unstable against the formation of a constant chromomagnetic field~\cite{Savvidy:1977as}, which suggests that $\langle G_{\mu\nu}^a G_{\mu\nu}^a\rangle > 0$. Although the gluon condensate cannot be visualized directly, it is known to have structure at intermediate scales consisting of monopoles~\cite{Shiba:1994db} and center vortices~\cite{Engelhardt:1999wr}.

The way we approach confinement is by systematically changing the scale at which the physical system is viewed. This is made possible by the gradient flow, which is a particular, infinitesimal realization of the coarse-graining step of momentum space renormalization group transformations. For a review and references see~\cite{Sonoda:2025tyu}. The energy density of the flowed field, $\langle G_{\mu\nu}^a G_{\mu\nu}^a\rangle$, is used to determine the running coupling $\alpha_{GF}(\mu)$ at flow times $t = 1/8\mu^2$, ranging from $t = 0$ to $t \approx 0.2 \,L^2$, thereby bridging the ultraviolet and infrared regimes. The result is that at long distances the coupling becomes infinitely strong, $\alpha_S \simeq \Lambda_S^2/\mu^2$, in accord with infrared slavery, the flip side of asymptotic freedom. This confines quarks and gluons to color-neutral composite states known as hadrons. It has been shown to give rise to a linearly rising static potential that satisfies Wilson's confinement criterion.

Our work represents a significant change in approach to confinement. The new development is that the running coupling in the infrared has become tangible analytically. On this basis a rigorous proof of confinement should be feasible.

\section*{Acknowledgements}

I like to thank my colleagues of the QCDSF collaboration, especially Utku Can, Roger Horsley, Yoshifumi Nakamura and James Zanotti, for their help and leadership in generating the configurations on which the numerical analyses are based. 

\section*{Appendix}

We will show now that the gluon condensate $G$ is independent of $\mu$ for finite flow times $t=1/8\mu^2$. According to~\cite{Luscher:2010iy}, the effective action at flow time $t$ is given by
\begin{equation}
  S_G^{\,t} = S_G + \frac{4}{3a^2} \int_0^t dt \int d^4x\, G_{\mu\nu}^a G_{\mu\nu}^a \,.
  \label{effact}
\end{equation}
From (\ref{effact}) we derive the flow-time derivative of the vacuum energy density,
\begin{equation}
  a^2 \frac{\partial}{\partial t} \langle G_{\mu\nu}^a G_{\mu\nu}^a \rangle = - \frac{4}{3} \int d^4x \,\langle G_{\mu\nu}^a G_{\mu\nu}^a(0) \; G_{\mu\nu}^a G_{\mu\nu}^a(x)\rangle_c \,.
  \label{corr}
\end{equation}
Here the subscript $c$ denotes the connected correlation function. At flow times $t > 0$ the operator $G_{\mu\nu}^a G_{\mu\nu}^a$ is related to the renormalized action density by a multiplicative factor that depends on the renormalization scale, in general $\mathcal{O}_R = Z(\mu)\, \mathcal{O}$, leading to the identification of an anomalous dimension. The result of the differential equation (\ref{corr}) is
\begin{equation}
  \langle G_{\mu\nu}^a G_{\mu\nu}^a \rangle \propto 1/t \;\; \therefore \;\; G \propto t^2 \langle  G_{\mu\nu}^a G_{\mu\nu}^a \rangle^2 = \textrm{constant} \,,
\end{equation}
aside from a constant, noninteracting value that applies to the large flow time limit $\langle E \rangle \simeq 8\pi \langle |Q|\rangle /V$ (Fig~\ref{fig1}). (Similarly, $\langle 0|G_{\mu\nu}^a G_{\mu\nu}^a |\sigma\rangle \propto 1/t$. See below.) 

An independent solution can be derived by evaluating the path integral (\ref{corr}). When field operators are averaged over finite regions of space-time, the central limit theorem (CLT) applies. The CLT states that the cumulative effect of many independent or weakly correlated fluctuations across a region tends to a Gaussian distribution. A Gaussian distribution of the vacuum energy consistently yields a $t$-dependence of the form $\langle E^2(t) \rangle \propto \langle E(t) \rangle^2 \propto 1/t^2$.

The original proof~\cite{Shifman:1978bx,Novikov:1979va} by sum rules considers the time ordered product of two conserved electromagnetic currents, 
\begin{equation}
  \Pi_{\mu\nu}(Q^2) = i \int d^4x\, e^{i\, q x}\, \langle 0|\textrm{T} [J_\mu(x)\; J_\nu(0)]|0\rangle = (q_\mu q_\nu - q^2)\, \Pi(Q^2) \,, \;\; Q^2 = -q^2 \,,
\end{equation}
where \vspace*{-0.25cm}
\begin{equation}
  J_\mu = \frac{1}{2} \left(\bar{u}\gamma_\mu u - \bar{d}\gamma_\mu d\right).
\end{equation}
The OPE of $\Pi(Q^2)$ can be used to define gluon and quark condensates~\cite{Shifman:1978bx,Gockeler:2000kj}. For mass-degenerate $u$ and $d$ quarks we have
\begin{equation}
  \Pi(Q^2) = c_0(Q^2)\, \langle \mathbbm{1}\rangle + c_4^G(Q^2)\, \frac{(\alpha_S/\pi)\, \langle G_{\mu\nu}^a G_{\mu\nu}^a\rangle}{24\, Q^4} + c_4^\Sigma(Q^2)\, \frac{m\, \langle \bar{u}u + \bar{d}d\rangle}{2\, Q^4} + O\left(1/Q^6\right) \,,
\end{equation}
where $c_0$, $c_4^G$ and $c_4^\Sigma$ are the Wilson coefficients. We adopt a suitable scheme $S$ to define renormalized Wilson coefficients and condensates. The scale parameter is conveniently taken to be $\mu^2=Q^2$. The Wilson coefficients $c_4^G$ and $c_4^\Sigma$ are of the form $c_4^{G,\Sigma} = 1 + O(\alpha_S)$. Apart from mixtures with the quark condensate, the gluon condensate is uniquely determined if the Wilson coefficient $c_0(Q^2)$ is free of power corrections of $O(1/Q^4)$. In this case $(\alpha_S/\pi)\, \langle G_{\mu\nu}^a G_{\mu\nu}^a\rangle$ will be independent of the renormalization scale. Alternatively, we could have considered purely gluonic currents~\cite{Narison:1984hu}, which are less familiar though. Power corrections can arise from renormalon contributions, which originate from the divergent behavior of the perturbative series, specifically in calculations involving loop momenta. We are interested in the gluon condensate at finite flow time. The gradient flow provides a way to regularize the theory by introducing a physical momentum scale that alters the behavior of the perturbative series. The flow time acts as an ultraviolet cut-off, thus eliminating the renormalon ambiguity. An explicit calculation can be found in~\cite{Beneke:2025hlg}. This eliminates infrared renormalon contributions as well~\cite{Novikov:1984rf}.

The connected correlation function on the right-hand side of (\ref{corr}) measures local excitations above the vacuum, which require finite energy. An important point to notice is that the derivative on the left-hand side is nonzero and finite, which suggests that the theory has a mass gap. By mass gap we understand the minimum positive energy of the lightest particle. The emphasis is on particle, as opposed to an effective scale possibly provided by the gluon mass in Landau gauge. The correlation function can be re-expressed using the spectral representation by
\begin{equation}
  - \frac{4}{3} \sum_n \int dt\; \big|\langle 0| G_{\mu\nu}^a  G_{\mu\nu}^a|n\rangle\big|^2 e^{- m_n t}  \,,
  \label{corr2}
\end{equation}
where $t$ is the Euclidian time, which should not be confused with the flow time. The operator $G_{\mu\nu}^a G_{\mu\nu}^a$ projects onto $J^{PC} = 0^{++}$ glueballs. We call the lightest glueball state $\sigma$, which is normalized to $|\sigma\rangle = \sqrt{2 m_\sigma} |1\rangle$. Evaluating the time integral in (\ref{corr2}) then gives
\begin{equation}
   - \frac{4}{3}\, \frac{\big|\langle 0| G_{\mu\nu}^a  G_{\mu\nu}^a|\sigma\rangle\big|^2}{2 m_\sigma^2} \,.
  \label{corr3}
\end{equation}
The derivative on the left-hand side of (\ref{corr}) can be worked out by expressing the flow time in terms of the running coupling. The final result is
\begin{equation} 
  \frac{\big|\langle 0| \left(\alpha_{GF}/\pi\right) \,G_{\mu\nu}^a  G_{\mu\nu}^a|\sigma\rangle\big|^2}{2 m_\sigma^2} \, = \, \frac{288}{\pi^3} \, \Lambda_{GF}^6\,,
  \label{corr4}
\end{equation}
where the operator $G_{\mu\nu}^a G_{\mu\nu}^a$ is replaced by the scale invariant operator $(\alpha_{GF}/\pi) \,G_{\mu\nu}^a G_{\mu\nu}^a$ in the process. Zero glueball mass would mean that the right-hand side of (\ref{corr}) tends to infinity at infinite times, which is excluded by the left-hand side being finite. No fluctuations above the vacuum at all would render the right-hand side zero, which is equally excluded.

%= |\sigmaowestThe operator $G_{\mu\nu}^a G_{\mu\nu}^a$ projects onto $0^{++}$ glueball states. The states are normalized as $\langle n|m\rangle = \delta_{nm}$. If there is a massless state $m_0 = 0$ (i.e. no mass gap), then the integral in (\ref{corr2}) will diverge. This contradicts the assumption of a finite gluon condensate.

%In the pure SU(3) gauge theory the nonperturbative vacuum is characterized by the gluon condensate $G$ and the topological susceptibility $\chi_t$. The topological susceptibility is known to be scale invariant as well~\cite{Nakamura:2021meh,ML},
%\begin{equation}
%   \frac{\partial}{\partial t} \, \chi_t = 0 \,.
%\end{equation}
%A nonvanishing topological susceptibility $\chi_t$ is inconceivable without a nonvanishing gluon condensate $G$. %This is what we would expect from a renormalization group transformation. 

\end{document}